\documentclass[11pt,twoside]{scrartcl}
\usepackage[top=3cm,bottom=2.5cm,inner=3.5cm,outer=2cm]{geometry}                
\usepackage[parfill]{parskip}    
\usepackage{graphicx}
\usepackage{amssymb}
\usepackage{epstopdf}
\usepackage{hyperref}
\usepackage[authoryear]{natbib}
\usepackage{fancyhdr}
\usepackage{framed} 
\usepackage{multicol} 
\usepackage{float} 
\usepackage{amsmath,amssymb}
\newcommand{\shorttitle}{Bispectra of Internal Tides}
\newcommand{\mytitle}{Bispectra of Internal Tides and Parametric Subharmonic
Instability}
\newcommand{\shortauthor}{E. Frajka-Williams et al.}
\newcommand{\myauthor}{Eleanor Frajka Williams, Eric L. Kunze, Jennifer A. MacKinnon}

\begin{document}
\thispagestyle{plain}
\section*{\LARGE\mytitle}
\subsection*{\myauthor}

\begin{framed}
\textit{This manuscript is based on the MSc work of EFW at the University of Washington.  It was submitted to JMR in June 2006.}
\end{framed}
\vspace{.2in}
\begin{abstract}
Bispectral analysis of the nonlinear resonant interaction known as parametric subharmonic instability (PSI) for a coherence semidiurnal internal tide demonstrates the ability of the bispectrum to identify and quantify the transfer rate.  Assuming that the interaction is confined to a vertical plane, energy equations transform in such a way that nonlinear terms become the third-moment spectral quantity known as the bispectrum. Bispectral transfer rates computed on PSI in an idealized, fully-nonlinear, non-hydrostatic Boussinesq model compare well to model growth rates of daughter waves. Bispectra also identify the nonlinear terms responsible for energy transfer.  Using resonance conditions for an M$_2$   tide, the locus of PSI wavenumber triads is determined as a function of parent-wave frequency and wavenumbers, latitude and range of daughter-wave frequencies. The locus is used to determine the expected bispectral signal of PSI in wavenumber space. Bispectra computed using velocity profiles from the HOME experiment are relatively noisy and the signal inconclusive.
\end{abstract}

\pagestyle{fancy}

%
%

%

%
%

\section{Introduction}
Away from sources, the energy spectrum of ocean internal waves is well-described by the Garrett- Munk model, a red spectrum with $-2$ slope in vertical wavenumber space \citep{GM79,Sherman-Pinkel-1991,Lvov-etal-2004}. Energy is added at large scales by tides and
wind-stress. Energy is lost to turbulence at O(1 m) scales due to wave breaking. Since vertical mixing plays an important role in the ocean overturning circulation \citep{Munk-Wunsch-1998}, oceanographers have long sought to understand the mechanisms responsible for the cascade of energy from large to small scales.

One likely mechanism, parametric subharmonic instability (PSI), is a resonant wave-triad interaction characterized by transfer of energy from a parent wave to two daughter waves of half-frequency and higher wavenumber. During the 1960s--80s, studies of PSI in a randomly-phased internal wave field led to construction of analytic solutions under small-amplitude assumptions. Numerical and laboratory experiments demonstrated PSI and identified conditions under which PSI occurs \citep{Muller-etal-1986}. Field attempts to identify PSI in the Arctic Ocean appeared successful but it was later determined that the apparent signal of PSI was due kinematic contamination of an Eulerian time series \citep{Neshyba-Sobey-1975,McComas-Briscoe-1980}. Later studies
concluded that the timescales of random-wave PSI interaction are too long to be physically relevant, and the bispectral signal too small to be measured \citep{McComas-1978,McComas-Briscoe-1980,Olbers-1976}. The thread of study was dropped.
More recently, \citet{Hibiya-etal-1996} used a two-dimensional numerical model to simulate in- ternal wave interactions, and postulated a role for PSI. Further numerical studies supported this conclusion, demonstrating that rapid energy transfer from an initial spike of energy occurs only if the spike satisfies frequency requirements for a PSI parent wave \citep{Hibiya-etal-1998,Hibiya-etal-1999,Hibiya-etal-2002}. Observations of diffusivity were not inconsistent with an expected latitudinal dependence of PSI for an M  semidiurnal tide \citep{Nagasawa-etal-2002}. Recent data and simulations associated with the Hawaiian Ocean Mixing Experiment (HOME) suggest that PSI of a coherent M  tide may be measurable in the ocean \citep{Rainville-2004,Carter-Gregg-2006,MacKinnon-Winters-2005}.

In this paper, the bispectrum is demonstrated to be an effective tool for identifying PSI in a
coherent internal tide, as well as determining the rate of transfer via PSI from the tide. In \S2, the fundamentals of wave-triad interaction theory and the specific characteristics PSI are laid out. In
\S3, triad resonance conditions are used to determine the locus of a PSI triad in vertical wavenumber space. In \S4, definitions and statistics of bispectra and bicoherences are given. In \S5, bispectral estimation of energy transfer rates are described. In \S6, these tools are applied to a numerical simulation of a mode-1 internal tide which undergoes PSI. Bispectrally computed rates of energy transfer compare favorably with energy content changes of the subharmonic during the spinup phase of a model run. Nonlinear terms responsible for the majority of energy transfer are identified and described. Bispectra of ocean data are shown in \S7. and the behavior of bispectra in the presence of complicating factors described. In \S8, we conclude with a few remarks concerning the applicability of these techniques to the ocean or atmosphere, and present a few caveats of these techniques.

\section{Parametric Subharmonic Instability of the M$_2$ Tide}

Resonant wave-triad interactions are nonlinear interactions between three freely-propagating internal waves. The fundamental interaction as described in \citet{Gill-1982} is that of two plane waves with wavevectors $\mathbf{k'}$ and $\mathbf{k''}$, and frequencies $\omega(\mathbf{k'})$ and $\omega(\mathbf{k''})$ determined by the dispersion relation.  The nonlinear terms in the equations of motion, $u\partial u/\partial x$, etc, appear as $exp\left\{i(\mathbf{k'}+\mathbf{k''})\mathbf{x}-i\left[\omega(\mathbf{k'})+\omega(\mathbf{k''})\right]t\right\}$, 
which is a forced-wave response with wavevector $\mathbf{k'}+\mathbf{k''}$ and frequency $\omega(\mathbf{k'})+\omega(\mathbf{k''})$, unless
\begin{equation}
\omega(\mathbf{k'}+\mathbf{k''}) = \omega(\mathbf{k'})+\mathbf{k''})
\end{equation}
If (1) is satisfied, the interaction is resonant and the nonlinear terms contribute to a freely propagating wave. Energy may transfer freely to or from any wave in a triad interaction, though net energy transfer may result depending on external sources and sinks. Since dissipation tends to occur at higher wavenumbers, energy preferentially transfers from lower to higher wavenumbers. Energy transfer rates depend on energy levels in triad waves with more energetic waves result in higher rates.

Three classes of resonant wave-triad interactions are described in \citet{McComas-Bretherton-1977}, distinguished by wavenumber and frequency. One such triad is parametric subharmonic  instability (PSI), so-called because half-frequency waves are produced. The forced pendulum is the classic example in which the restoring force gravity is modified at twice the natural frequency of the pendulum.  In the case of ocean internal waves, \citet{Muller-etal-1986} claim that energy is transferred due modification of the buoyancy frequency of small-scale waves by the large scale wave ($<w\partial b/\partial z>$).  \citet{McEwan-Robinson-1975} identified the mechanism for energy transfer as a rotating of isopycnals in a cylindrical tank ($<u\partial b/\partial x>$).  The physical mechanism will be examined further in \S6.
For plane waves with phase $\theta = \mathbf{k}\cdot\mathbf{x}-\omega\cdot t$, equation (1) may be separated into the diagnostic resonance conditions,
\begin{align}
\mathbf{k}+\mathbf{k'}+\mathbf{k''}&=0\\
\omega+\omega'+\omega''&=0\ .
\end{align}

Waves must also satisfy the dispersion relation for internal waves
\begin{equation}
\frac{k_1^2+k_2^2}{k_3^2} = \frac{\omega^2-f^2}{N^2-\omega^2}
\end{equation}
where $\mathbf{k}=(k_1,k_2,k_3)$ is the wave vector in $(x,y,z)$ directions, and the frequency of a freely-propagating internal wave is confined between the buoyancy frequency $N$  and Coriolis frequency $f$,
\begin{equation*}
f<\omega<N\ .
\end{equation*}
Since $f=2\Omega\sin\phi$ depends on latitude $\phi$, and the parent wave must have frequency of twice the daughter wave, this last condition translates into a latitudinal constraint on PSI for the M$_2$  (12.4 hr) semidiurnal tide. Where M$_2$  is the parent wave, PSI may only occur equatorward of 28.9$^\circ$  latitude (where the M$_2$ frequency is exactly twice the Coriolis frequency). See Figure 1.

Waves not satisfying resonance condition (1) are forced (non-propagating), and waves not satisfying the frequency restrictions are trapped, with an exponential rather than oscillating solution. Both of these types of interactions are confined to the generation location and will not be considered here.

\section{Wavenumber Triad  Ellipses}

Assuming the parent-wave frequency and wavenumber are known, and that the strongest transfers are confined to a vertical plane ($k_2=0$  for all waves), analytical expressions for the frequencies and wavenumbers of PSI daughter waves are derived.  Fundamental parameters are latitude and
$\epsilon$, a measure of how far the daughter-wave frequencies (denoted $\omega'$  and $\omega''$) deviate from half the parent-wave frequency  $\omega/2$,
\begin{equation}
\epsilon \equiv \left|\frac{\omega'}{\omega}-\frac{1}{2}\right| = \left|\frac{\omega''}{\omega}-\frac{1}{2}\right|\ .
\end{equation}
Without loss of generality, we can say $\epsilon=\omega'/\omega-1/2=1/2-\omega''/\omega$.  Nondimensionalizing frequencies and wavenumbers by the parent-wave frequency $\omega$ and vertical wavenumber $k_3$, the resonance conditions (2) and (3) can be solved for the ratio of the daughter- to parent-wave's vertical wavenumber,
\begin{equation}
\frac{k_3'}{k_3}=\frac{\sqrt{(1-\omega'/\omega)^2-(f/\omega)^2}\pm\sqrt{1-(f/\omega)^2}}{\sqrt{(\omega'/\omega)^2-(f/\omega)^2}\pm\sqrt{(1-\omega'/\omega)^2-(f/\omega)^2}}\ .
\end{equation}
Substituting in $\epsilon$ for $\omega'$, we have
\begin{equation}
\frac{k_3'}{k_3}=\frac{\sqrt{(1-\epsilon)^2-(f/\omega)^2}\pm\sqrt{1-(f/\omega)^2}}{\sqrt{(1/2+\epsilon)^2-(f/\omega)^2}\pm\sqrt{(1/2-\epsilon)^2-(f/\omega)^2}}\ .
\end{equation}
Two sign choices result from the roots. Solutions are plotted in Figure 2.

Assuming the daughter waves have exactly half the parent-wave frequency ($\epsilon=0$), that is, they are exactly subharmonic, then minus sign choice in (7) yields an infinite daughter-wave vertical wavenumber which is aphysical.  The solution of (7) for the plus sign choice of PSI of the M$_2$ internal tide is shown in Figure 3 as a function of latitude.

Energy transfer via resonant triads works for any triad satisfying the resonance conditions.
The PSI class of triads appears to be particularly efficient, but is not limited to energy transfer to exactly half the parent-wave frequency. Slight deviations from exactly half frequency create large variations in the expected daughter-wave vertical wavenumbers. For small but non-zero $\epsilon$, the locus of points which are solutions to (7) is given by the ellipse
\begin{equation}
K_1^2+K_1K_2+K_2^2 = \left(\frac{k_3'}{k_3}\right)^2 + \frac{k_3'}{k_3}+1\ ,
\end{equation}
where  $K_1$ and $K_2$ are vertical wavenumbers in a triad of either parent or daughter waves, nondimensionalized by the parent wave.  The third wave of the triad is $K_3=-K_1-K_2$.  Due to nondimensionalization, one of the three of $K_1$, $K_2$ and $K_3$  is 1. The ratio $k_3'/k_3$ and the size of the wavenumber ellipse increase as  $\epsilon$ decreases (Figure 4), representing a transfer of energy to higher wavenumbers.

\section{Statistical Techniques}
In this section, we review the definitions, use and interpretation of the bispectrum and bicoherence. The bispectrum is the third-order moment equivalent of the power spectrum. Just as power spectra decompose variance in a signal into individual frequency or wavenumber bands, bispectra express second-order interactions between frequency or wavenumber components by decomposing skew- ness into spectral bands.

\subsection{Bispectrum}
For a continuous real time-series of a zero-mean perturbation quantity, e.g., horizontal perturbation velocity,	 $u(t)$, the Fourier coefficients are
\begin{equation}
U(\omega)=\frac{1}{M}\sum^T_{t=0}u(t)e^{-i\omega t}\ ,
\end{equation}
where time $t=m\delta t$ with $1\leq m\leq M$  and $\delta t=T/M$ the sampling resolution. Frequency is given by $\omega=j\delta\omega$ where  $-M/2\leq j\leq M/2$, the lowest resolved frequency is  $\delta\omega=2\pi/T$ and the highest resolved (Nyquist) frequency is $\omega_{max}=2\pi/\delta t$. The power spectrum of $u$ is defined as
\begin{equation*}
S_{uu}(\omega)\,d\omega = U(\omega)U(-\omega)\ ,
\end{equation*}
which is the decomposition of second-moment variance into frequency bands,
\begin{equation}
\sum_{\omega=-\omega_{max}}^{\omega_{max}} S_{uu}(\omega)\,d\omega = \frac{1}{M}\sum_{m=1}^Mu(t)^2=var(u)\ .
\end{equation}
This definition of the spectrum is equivalent to $S_{uu}(\omega)\,d\omega=U(\omega)U^*(\omega)$ for a real time-series with $(\cdot)^*$ the complex conjugate, but more clearly reveals the relationship between resonance conditions
and the bispectrum. The third central moment of a time-series is skewness,
\begin{equation}
skew(u)\equiv\frac{1}{M}\sum_{m=-0}^Mu(t)^3=\sum_{\omega'=-\omega_{max}}^{\omega_{max}} \sum_{\omega''=-\omega_{max}}^{\omega_{max}} U(\omega')U(\omega'')U(-\omega'-\omega'')
\end{equation}
which leads to the definition of the auto-bispectrum, or bispectrum of a signal with itself,
\begin{equation}
BiS_{uuu}(\omega',\omega'')\,d\omega'\,d\omega'' \equiv U(\omega')U(\omega'')U(-\omega'-\omega'')\ .
\end{equation}
By construction, a nonzero bispectrum at the frequency pair $(\omega',\omega'')$ identifies an energetic wave triad $(\omega',\omega'',\omega)$ which satisfies frequency resonance where the third member of the triad $\omega=-\omega'-\omega''$.  However, the bispectrum does not constrain triads to satisfy the dispersion relation. For PSI, the expected frequency bispectral signal is large at frequency pairs $BiS(\omega/2,\omega/2)$ or $BiS(\omega/2,-\omega)$ 
where the third member of the triad identified by the bispectrum is $\omega=-(\omega/2+\omega/2)$ or $\omega/2=-(\omega/2-\omega)$, respectively.

In practice, the bispectrum is estimated by averaging over many realizations,
\begin{equation}
\widehat{BiS}_{uuu}(\omega',\omega'')\,d\omega'\,d\omega'' = E\left[U(\omega')U(\omega'')U(-\omega'-\omega'')\right]
\end{equation}
where $E[\cdot]$ is the expectation. Using polar form, the bispectrum
\begin{equation}
\widehat{BiS}_{uuu}(\omega',\omega'')\,d\omega'\,d\omega'' = E\left[|U(\omega')||U(\omega'')||U(-\omega'-\omega'')|e^{i(\theta(\omega')+\theta(\omega'')+\theta(-\omega'-\omega''))}\right]
\end{equation}
is clearly a function of both the power of individual triad waves and phase-locking between triad waves. This means that, if the power in individual waves satisfying only (3) is large and the number of realizations few, then the bispectrum averaged over these realizations may be nonzero even though the degree of phase-locking is low. The result is a false signature wave-triad interaction. The dependence of the bispectrum on the power in the spectrum motivates the construction of the bicoherence.

\subsection{Bicoherence and Significance}

The bicoherence is the normalized magnitude of the bispectrum, showing only the degree of phase- locking between $\omega$    components. The bicoherence
\begin{equation}
\hat{b}^2_{uuu}(\omega',\omega'')\,d\omega'\,d\omega'' \equiv \frac{\left| \widehat{BiS}_{uuu}(\omega',\omega'')\,d\omega'\,d\omega''\right|^2}{E\left[\left|U(\omega')U(\omega'')\right|^2\right]E\left[\left|U(-\omega'-\omega'')\right|^2\right]}\ ,
\end{equation}
is bound between 0 and 1. The bicoherence of one realization (no averaging) is identically 1; the bicoherence of an infinite number of realizations with random phase is identically 0.  If a single wave-triad interaction dominates the power spectrum at each of the constituent triad frequencies, then the bicoherence at that triad will be significant.

\citet{Elgar-Guza-1988} calculated significance levels for zero bicoherence using Monte Carlo methods and find that, in agreement with theoretical levels \citep{Haubrich-1965}, data are 99\% significant with a bicoherence above $\sqrt{9.2/ndof}$ where  $ndof$ is the number of degrees of freedom, twice the number of independent realizations. Other levels are   $95\%=\sqrt{6/ndof}$, $90\%=\sqrt{4.6/ndof}$ and $80\%=\sqrt{3.2/ndof}$. In the ocean, an energetic background continuum of randomly-phased, non-interacting internal waves with only a finite number of available realizations tends to lower the bicoherence.

\subsection{Multidimensional Bispectra}

For multiple independent variables $(x,y,z,t)$, wavenumber-frequency Fourier coefficients are defined
\begin{equation*}
U(\mathbf{k},\omega)=\frac{1}{M}\frac{1}{\mathbf{P}}\sum_{t=0}^T\sum_{x=0}^\mathbf{L}u(\mathbf{k},t)e^{i2\pi(\mathbf{k}\cdot\mathbf{x}-\omega t)}\ ,
\end{equation*}
where space vector  $\mathbf{x}=(x,y,z)=p\,\delta x$ with  $1\leq p\leq P$, $\delta x=\mathbf{L}/P$ the sampling resolution. (To simplify notation, the number of samples in   $x$-, $y$- and  $z$- dimensions has been assumed the same, but the definition is easily modified if not.) The four-dimensional power spectrum is
\begin{equation*}
\widehat{S}_{uu}(\mathbf{k},\omega)\,d\mathbf{k}\,d\omega = E\left[U(\mathbf{k},\omega)U(-\mathbf{k},-\omega)\right]\ ,
\end{equation*}
and the eight-dimensional bispectrum is,
 \begin{equation*}
\widehat{BiS}_{uuu}(\mathbf{k'},\omega',\mathbf{k''},\omega'')\,d\mathbf{k'}\,d\omega'\,d\mathbf{k''}\,d\omega'' = E\left[U(\mathbf{k'},\omega')U(\mathbf{k''},\omega'')U(-\mathbf{k'}-\mathbf{k''},-\omega'-\omega'')\right]
\end{equation*}
Lower-dimensional bispectra can be computed as integrals of higher-dimensional bispectra over the omitted dimensions, for example,
\begin{equation}
{BiS}_{uuu}(k_3',\omega',k_3'',\omega'')\,dk_3'\,d\omega'\,dk_3''\,d\omega'' = \sum_{k_1'}\sum_{k_1''}\sum_{k_2'}\sum_{k_2''}{BiS}_{uuu}(\mathbf{k'},\omega',\mathbf{k''},\omega'')\,d\mathbf{k'}\,d\omega'\,d\mathbf{k''}\,d\omega''
\end{equation}
and likewise,
\begin{equation*}
{BiS}_{uuu}(\omega',\omega'')\,d\omega'\,d\omega'' = \sum_{k_3'}\sum_{k_3''}{BiS}_{uuu}(k_3',\omega',k_3'',\omega'')\,dk_3'\,d\omega'\,dk_3''\,d\omega''\ .
\end{equation*}

To produce a two-dimensional frequency bispectrum, the bispectral peak at a given triad of wavevectors and frequencies in the eight-dimensional bispectrum is averaged over all wavevectors. This averaging obscures the signal of parametric subharmonic instability for a single energetic triad in the absence of multi-dimensional data. Bicoherence, the normalized bispectrum, also gives better (higher) estimates for phase-locking in multiple dimensions because it is not averaged over non-PSI triads in the omitted dimension.  (Unlike the bispectrum, the two-dimensional bicoherence is not the integral of the four-dimensional bicoherence over the omitted dimension since the integral of the four-dimensional bicoherence is an integral over a quantity with absolute values, $\sum|a|\neq\left|\sum a\right|$.)

\subsection{Cross-bispectrum}

The cross-bispectrum between two different oceanic variables is defined
\begin{equation}
 \widehat{BiS}_{uvw}(\omega',\omega'')\,d\omega'\,d\omega'' \equiv E\left[U(\omega')V(\omega'')W(-\omega'-\omega'')\right]
 \end{equation}
where the order of $U$, $V$ and $W$ is nontrivial.  Cross-bispectra are necessary to quantify energy transfer rates, described next in \S5.

\section{Transfer of Energy}
As shown in the previous section, the bispectrum identifies wave triads by construction, showing phase-locking between energetic triads of waves which satisfy the resonance conditions.  Given enough dimensions of data, it can be used to quantify the rate of nonlinear transfer of energy via a wave-triad interaction. This is demonstrated, following the unpublished manuscript of \citet{McComas-1978}. Starting with the inviscid equations of motion,
\begin{align}
\frac{\partial\mathbf{u}}{\partial t}+\mathbf{u}\cdot\nabla\mathbf{u}-f\times \mathbf{u}+\nabla\cdot p=0\\
\frac{\partial b}{\partial t}+\mathbf{u}\cdot\nabla{b} =-N^2w\ ,
\end{align}
then assuming we have a triad interaction so that $u_{tot}=u(t)+e^{-i\omega t}+u'(t)e^{-i\omega't}+u''(t)e^{-i\omega''t}$, where
$u$, $u'$ and $u''$ are the velocity amplitudes associated with three waves with frequencies $\omega$, $\omega'$ and $\omega''$.
 Separating each of the wave amplitudes into $u(t)=u_0+u_1(t)$ where $\partial u_0/\partial t$ is small, then the linear equations are
\begin{align*}
\frac{\partial \mathbf{u_{lin}}}{\partial t}-2\Omega\times\mathbf{u_{lin}}+\nabla\cdot p=0\ ,\\
\frac{\partial b_{lin}}{\partial t} = -N^2\omega_{lin}\ ,
\end{align*}
where $u_{lin}=u_0e^{-i\omega t}+u'_0e^{-i\omega't}+u_0''e^{-i\omega''t}$, and nonlinear amplitude growth equations,
\begin{align}
\frac{\partial \mathbf{u_{nl}}}{\partial t} + \mathbf{u_{nl}}\cdot\nabla\mathbf{u_{nl}} = 0\ ,\\
\frac{\partial b_{nl}}{\partial t} + \mathbf{u_{nl}}\cdot\nabla b_{nl}=0\ ,
\end{align}
where $u_{nl}=u_1(t)e^{-i\omega t}+u_1'(t)e^{-i\omega't}+u_1''(t)e^{-i\omega''t}$. Expanding the nonlinear, real   $u$-momentum equation and dotting by  $u_{nl}^*$, the complex conjugate of $u_{nl}$, gives 117 terms in the  $u$-momentum equation, including
\begin{equation}
\frac{1}{2}\frac{\partial u_1^2}{\partial t} + u_1u_1'\frac{\partial u_1''}{\partial x}e^{i(\omega-\omega'-\omega'')t}+\ldots\ .
\end{equation}
If   $\omega=\omega'+\omega''$, that is, if the three waves satisfy the frequency resonance condition, then those terms with frequencies summing to 0 contribute to growth (or decay) of the wave amplitudes. The others do not result in energy transfer, when averaged over a wave period. This equation (22), and multiplying (21) by vertical displacement $\zeta_{nl}^*=-b_{nl}^*/N^2$, form the kinetic and potential energy equations. Fourier-transforming into frequency and wavenumber space, the energy evolution equations become
\begin{align}
\frac{1}{2}&\frac{\partial U(\mathbf{k},\omega)U(-\mathbf{k},-\omega)}{\partial t}\\\notag
&=-i\sum_{\omega'}\sum_{\omega''}\sum_{\mathbf{k'}}\sum_{\mathbf{k''}}\left\{k_j'U(\mathbf{k'},\omega')U_j(\mathbf{k''},\omega'')U(\mathbf{k},\omega)+k_j''U_j(\mathbf{k'},\omega)U(\mathbf{k''},\omega'')U(\mathbf{k},\omega)\right\}\\\notag
&\qquad\delta(\mathbf{k}+\mathbf{k'}+\mathbf{k''})\,\delta(\omega+\omega'+\omega'')\ ,\\
\frac{1}{2}&\frac{\partial V(\mathbf{k},\omega)V(-\mathbf{k},-\omega)}{\partial t}\\\notag
&=-i\sum_{\omega'}\sum_{\omega''}\sum_{\mathbf{k'}}\sum_{\mathbf{k''}}\left\{k_j'V(\mathbf{k'},\omega')U_j(\mathbf{k''},\omega'')V(\mathbf{k},\omega)+k_j''U_j(\mathbf{k'},\omega)V(\mathbf{k''},\omega'')V(\mathbf{k},\omega)\right\}\\\notag
&\qquad\delta(\mathbf{k}+\mathbf{k'}+\mathbf{k''})\,\delta(\omega+\omega'+\omega'')\ ,\\
\frac{1}{2N^2}&\frac{\partial B(\mathbf{k},\omega)B(-\mathbf{k},-\omega)}{\partial t}\\\notag
&=-i\sum_{\omega'}\sum_{\omega''}\sum_{\mathbf{k'}}\sum_{\mathbf{k''}}\left\{k_j'B(\mathbf{k'},\omega')U_j(\mathbf{k''},\omega'')B(\mathbf{k},\omega)+k_j''U_j(\mathbf{k'},\omega)B(\mathbf{k''},\omega'')B(\mathbf{k},\omega)\right\}\\\notag
&\qquad\delta(\mathbf{k}+\mathbf{k'}+\mathbf{k''})\,\delta(\omega+\omega'+\omega'')\ ,
\end{align}
showing growth of a wave at $(\mathbf{k},\omega)$ via nonlinear terms.  We have neglected the $W^2$	equation assuming that $W^2\ll U^2$. The delta functions ensure that only frequencies (wavenumbers)	 $\omega'$ and $\omega''$ ($k'$ and $k''$) which sum to the frequency (wavenumber) on the LHS, $\omega=\omega'-\omega''$ ($k=-k'-k''$) are included. Rewritten in terms of the energy spectrum and multi-dimensional cross-bispectra, the rate of energy transfer is
\begin{align}
\frac{\partial E(\mathbf{k},\omega)\,d\mathbf{k}\,d\omega}{\partial t} = -i\sum_{\omega'}&\sum_{\omega''}\sum_{\mathbf{k'}}\sum_{\mathbf{k''}}\sum_j\\\notag
&\left[k_j'\left\{BiS_{u_iu_ju_i}(\mathbf{k'},\omega',\mathbf{k''},\omega'')+BiS_{bu_jb}(\mathbf{k'},\omega,\mathbf{k''},\omega'')\right\}\right.+\\\notag
&\ \left.k_j''\left\{BiS_{u_ju_iu_i}(\mathbf{k'},\omega',\mathbf{k''},\omega'')+BiS_{u_jb_ib_i}(\mathbf{k'},\omega',\mathbf{k''},\omega'')
\right\}\right]\\\notag
&\ \delta(\mathbf{k}+\mathbf{k'}+\mathbf{k''})\delta(\omega+\omega'+\omega'')\,d\mathbf{k'}\,d\mathbf{k''}\,d\omega'\,d\omega''\ ,
\end{align}
which we will call the ``transfer bispectrum''. The real part contributes to energy transfer and the imaginary part to changes phase on long timescales, causing the three waves to move in- and out- of-phase with each other as the interaction proceeds. The rate of energy transfer into a particular frequency or wavenumber band is thus proportional to both the magnitude of all three waves, and their degree of phase-locking, with additional dependence on the wavenumbers of the waves. All other things being equal, a transfer of energy to higher wavenumbers will proceed more rapidly than to lower wavenumbers.
The alternative method of calculating energy transfer via nonlinear wave-wave interactions would be to bandpass signals in time (space) then take the appropriate derivatives and multiply series together, as in \citet{MacKinnon-Winters-2005}, $v_{subharmonic}\times v_{subharmonic}\times \partial V_{M_2}/\partial y$. The bispectrum computes the same quantity, assuming plane waves, $ik_2=\partial/\partial y$ and at least two dimensions of data, e.g. $z$ and $t$, in order to infer the remaining wavenumbers $k_1'$, $k_1''$, $k_2'$  and $k_2''$ from the dispersion relation.

\section{Bispectra of Model Output}

A fully-nonlinear, three-dimensional, pseudo-spectral non-hydrostatic model \citep{Winters-etal-2004} was used to simulate PSI. The model has periodic side boundaries, sponge layers at the top and bottom of a 4-km deep, uniformly stratified ocean ($N=2\times 10^{-3}$ rad s$^{-1}$). Forced by a mode-1, M$_2$-frequency internal tide at 21$^\circ$ latitude with a small amount of noise, higher-wavenumber near half-M$_2$ frequency waves develop.  The eastward velocity profile time-series (Figure 5) clearly shows higher-wavenumber waves forming near the surface and bottom where mode-1 wave  $u$- velocities are largest.

\subsection{Spectral Estimation}

Frequency spectra were calculated in half-overlapping 5-day blocks. Vertical wavenumber spectra are more challenging to compute. In a variable stratification ocean, since perturbation quantities scale with stratification, the nonlinear terms in the energy equation and the energy transfer rates themselves depend strongly on stratification. Nevertheless, we will follow the standard procedure of computing stretched depth coordinates for the profiles, to transform the data onto an equivalent constant-stratification ocean.  The data are then normalized appropriately by the buoyancy frequency profile. The first mode is then a half-sine or cosine, so slight modifications to standard Fourier spectra must be made. The alternative is to create a set of normal modes for a given mean stratification profile. But since normal modes are really standing-mode solutions, there is no phase information associated with these modes and vertical wavenumber bispectra cannot be calculated.

Since the model has constant stratification, mode fits of this data are simply Fourier spectra
with the inclusion of the first modes. Figure 6 shows model two-dimensional power spectrum in frequency and vertical wavenumber.  Energy is apparent at mode-1 and the M$_2$ tidal frequency (the forcing in the model) as well as near mode-10 and half-M$_2$ frequency.  Since there is no forcing at mode-10 nor at half-M$_2$ frequency, one may conclude that the energy is transferred via nonlinear processes.  In a model, the amount of energy transferred to this secondary wave may be calculated by determining how much energy is gained by that wave, while adjusting for loss due to dissipation. The real ocean is more complicated, with multiple forcings at different scales, and energy propagating or being advected into or out of the region.  In the absence of detailed knowledge of sources and sinks, spectra do not quantify energy transfer through the spectrum. Bispectra do.

\subsection{Results: Frequency Bispectra}

Since PSI is characterized by a frequency relationship between parent and daughter waves, the surest way to identify it is through the frequency bispectrum, shown for the model data  $u$-perturbation velocity in Figure 7.  Axes are normalized by the M$_2$  frequency so that the signature of PSI is a peak at $(1/2,1/2)$  representing the triad satisfying the frequency resonance condition $\omega=-(0.5\omega_{M_2}+0.5\omega_{M_2})$. The peak at $(1, -1/2)$ is also PSI, corresponding to the same triad in a different order and called a difference interaction, since the resonance condition is $0.5\omega_{M_2}=-(\omega-0.5\omega_{M_2})$. Remaining peaks are due to symmetries in the bispectrum (Neshyba and Sobey, 1975). The independent region in the bispectrum and bicoherence is bound by   $|\omega''|\leq\omega'$.

Because it is a spectrally very clean, data never become significantly decorrelated, and so only
represent one independent realization. Though PSI is the most likely cause of energy transfer to half-frequency, significance tests do not apply.  The bicoherence only serves to indicate a high- degree of phase-locking at the triad waves, relative to other wave triads.

\subsection{Results: Wavenumber Bispectra}

The cascade of energy to turbulence is primarily concerned with vertical scales. Energy at smaller vertical scales dissipates much more quickly than energy at large vertical scales. Therefore, though the frequency bispectrum may easily identify PSI, the vertical wavenumber bispectrum is more relevant for our understanding the energy cascade and perhaps maintenance of the GM internal wave spectrum.

Ellipse (8) can be used to determine if a wavenumber bispectral signal is due to PSI. Vertical
wavenumber bispectrum and bicoherence of $u$-perturbation velocity data are shown in Figures 8a and b.  Axes are normalized by the parent-wave vertical wavenumber.  Thus, the peak at $(1,9)$ corresponds to a triad at modenumbers $(1,9,-10)$. Peaks lie on ellipse (8) corresponding to an $\epsilon=0.02$ and $k_2'=-k_2''=0$. The gap between peaks in daughter-wave frequencies independently verifies an	(Figure 6). This is however, the limit of $\epsilon$ resolution for a 25-day time series, $\delta\epsilon\approx 0.021$. Since the bicoherence also shows a high degree of phase-locking at these triads, we conclude that PSI is occurring.

Though these 6 peaks are solutions to an ellipse, there are other triads which are also solutions, including $(4,-5,1)$ which is not large in the bispectrum. One possible explanation for this is that, because the transfer of energy depends on wavenumber (26), triads with higher daughter wavenumbers have higher transfer rates for the same magnitude parent and daughter waves. One might also expect more energy in triads with higher wavenumbers still, except that energy is dissipated more rapidly at higher wavenumbers.

Though the energy source in the model is narrowband, peaks are diffuse in wavenumber space.
This may be due to induced diffusion, another wave-triad interaction identified by \citet{McComas-Bretherton-1977}, which dominates energy transfer at higher wavenumbers spreading or ``diffusing'' energy into nearby wavenumbers. Using the ray-tracing to calculate wavenumber evolution \citep{Gill-1982}, this effect can be quantified as
\begin{equation}
\frac{\delta k_3'}{k_3'} = \frac{k_1'}{k_3'\omega'}\frac{\partial u}{\partial z}\ ,
\end{equation} 
which is about $0.04$--$0.07$ spread in nondimensional wavenumber space around the ellipse for the model data. This is negligible.
 
\subsection{Multi-dimensional Bispectra}

One slice of the four-dimensional bispectrum of model data, $BiS_{uuu}(k_3',\omega',k_3'',\omega'')$ for $\omega'=0.5\omega_{M_2}$ and $\omega''=-\omega_{M_2}$  is shown in Figure 9. The horizontal axis is the vertical wavenumber corresponding to $\omega''$ and the vertical axis the vertical wavenumber corresponding to $\omega'$. Thus, the peaks at  $(\pm10,\pm1)$ and $(\pm11,\mp1)$                     correspond to the frequency triads, in order,
\begin{align*}
0.5\omega_{M_2}-\omega_{M_2}+0.5\omega_{M_2}&=0\\
10+1-11&=0\\
11-1-10&=0\\
-11+1+10&=0\\
-10-1+11&=0\ .
\end{align*}
The four-dimensional bicoherence shows high phase-locking at these triads as well as at sums of the daughter wavenumbers.  However, since these triads are not accompanied by corresponding peaks in the bispectrum, this phase-locking is not accompanied by peaks in the bispectrum, so though energy may be transferred, the magnitude is small and peaks represent triads of negligible physical importance.

\subsection{Results: Transfer of Energy}

To confirm that the transfer bispectrum accurately computes the energy transfer rate in the model, we compare energy in the daughter waves during the spinup phase of the model with the time integral of the energy transfer bispectrum into the daughter waves. Subharmonic energy and energy buildup were integrated over the  $x$-domain to smooth the signal.

Wavenumber coefficients of the transfer bispectrum are computed assuming $k_2=0$ as
\begin{equation}
\hat{k}_1' = \sqrt{\frac{k_3'^2(\omega'^2-f^2)}{N^2-\omega'^2}}\ .
\end{equation}
Given $x$-information from the model, $BiS_{uu_xu}(\omega',\omega'')$ is compared to $\sum_{k_3'}\sum_{k_3''}i\hat{k}_1''BiS_{uuu}(k_3',\omega',k_3'',\omega'')$
with good agreement, confirming (28). Integrated transfer bispectra summing over all nonlinear terms (blue) and subharmonic energy (red) during the spinup phase are shown in Figure 10. Subharmonic energy is calculated as
\begin{equation*}
E(0.5\omega_{M_2})=U(0.5\omega_{M_2})U(-0.5\omega_{M_2})+V(0.5\omega_{M_2})V(-0.5\omega_{M_2})+1/N^2\,B(0.5\omega_{M_2})B(-0.5\omega_{M_2})\ .
\end{equation*}
The blue curve is the time integral of energy transfer into the subharmonic by PSI, given by the right-hand side of (26) with  $\omega'=0.5\omega_{M_2}$ and $\omega''=-\omega_{M_2}$. Bispectra are estimated over blocks in time, in order to determine frequencies, which represents a running average of energy transferred by PSI. Disagreement between the curves is due to dissipation of energy in the subharmonic frequency not accounted for in the transfer bispectrum. Slopes and magnitudes of the two estimates of energy transfer show good agreement.
 
We also use the transfer bispectrum to determine the nonlinear term responsible for the majority of energy transfer to the subharmonic. Previous work by \citet{McEwan-Robinson-1975}, \citet{Hibiya-etal-2002} and \citet{Carter-Gregg-2006} claim that buoyancy modulation is responsible for energy transfer via PSI. However, upon computing the transfer bispectrum for 18 of the nonlinear terms (neglecting terms with $w^2$), we find that $u\partial u/\partial x$ dominates energy transfer. In the model, $u^2\gg v^2,w^2,b^2/N^2$.

The nonlinear terms in the energy equations are like periodic sources on the right-hand side of the equation. They do not change the dispersion relation of the waves, which is associated with the linear solution. Kinematically, the source terms for a daughter wave are just the product between the parent and the other daughter wave with
\begin{equation}
e^{-i\omega_{M_2}t}e^{-0.5i\omega_{M_2}t}=e^{-0.5i\omega_{M_2}t}\ .
\end{equation}

Near the turning latitude of 28.9$^\circ$, energy in the near-inertial daughter waves will be more concentrated in kinetic energy, suggesting that the $u\partial u/\partial x$ will dominate if the $u$-velocity is aligned with the parent-wave velocity. This may change closer to the equator or in areas of strong stratification, where parent- and daughter-wave energy may have more potential relative to kinetic energy. However, a latitudinal dependence is not observed in ratios of high-wavenumber available potential energy to horizontal kinetic energy computed from LADCP/CTD data \citep{Kunze-etal-2005}. 

\subsection{Rates of Energy transfer}

The rate of energy transfer via PSI is controlled by a number of factors. The energy evolution equations and the transfer bispectrum show that energy transfer increases with increasing amplitude, wavenumbers, and degree-of-phase-locking. The amplitude of triad waves depends on both the source and the sink. For the M$_2$  parent wave, the source is conversion from the surface to internal tide and sinks are net energy transfer to daughter waves via PSI and transfer to other smaller scale internal waves.  For daughter waves, the source is energy transfer from the parent wave and the sink is transfer to smaller scale internal waves. These transfers to smaller scale internal waves may be parameterized by the Gregg-Henyey dissipation parameterization which is based on wave/wave interaction theory.

Energy levels in the parent wave are governed by the balance between sources and sinks,
\begin{align}
\frac{dE_{M_2}}{dt} &= F-PSI_{M_2\rightarrow 0.5M_2}-\epsilon_{GH}(k_3)=0\\\notag
&=F-Re\left\{k_1'UU'U''+\ldots\right\}-\epsilon_0\frac{N^2}{N_0^2}\frac{\left<U_z^2\right>^2}{\left<U_{z,GM}^2\right>^2}f(R_\omega)\\\notag
&=F-Re\left\{k_1'UU'U''+\ldots\right\}-\alpha k_3^4U^4f(R_\omega)\ .\notag
\end{align}
where now the $\epsilon_{GH}$ represents dissipation rather than the parameter described in (5).
If the parent wave is mode-1, there will be little loss to dissipation so the third term may be neglected, leaving the balance,
\begin{equation}
\frac{dE_{M_2}}{dt} = F-Re\left\{k_1'UU'U''+\ldots\right\}\ .
\end{equation}
The parent-wave amplitude is $U$, $V$, \ldots, and depends on source energy, parent- and daughter-wave wavenumbers, and daughter-wave amplitudes.  Daughter-wave amplitudes are governed by the balance between PSI and dissipation,
\begin{align}
\frac{dE_{0.5M_2}}{dt} &= PSI_{M_2\rightarrow 0.5M_2}-\epsilon_{GH}(k_3')=0\\\notag
&=Re\left\{k_1UU'U''+\ldots\right\}-\epsilon_0\frac{N^2}{N_0^2}\frac{N^2}{N_0^2}\frac{\left<U_z'^2\right>^2}{\left<U_{z,GM}^2\right>^2}f(R_\omega')\\\notag
&=F-Re\left\{k_1UU'U''+\ldots\right\}-\alpha k_3'^4U'^4f(R_\omega')\ .\notag
\end{align}
The shear-to-strain ratio,
\begin{equation*}
R_{\omega'}=\frac{\left<U_z'^2\right>}{N^2\left<\zeta_z'^2\right>}\ ,
\end{equation*}
which depends on frequency $\omega$.

If the parent-wave frequency and wavenumber are known, source strength known, background GM levels for stratification and shear assumed, and if daughter waves have the same amplitude ($O(U')=O(U'')$) and wavenumber ($k_j'=k_j''$), then (31) and (32) may be solved for daughter wavenumber and amplitude. This would give some indication of what equilibrium energy levels would be for a given PSI interaction.

\section{Bispectra of Ocean Data}

The Hawaiian Ridge, a steep ridge lying nearly perpendicular to the barotropic M$_2$  tide in the Pacific, is a site of intense internal tide generation.  As a part of the Hawaiian Ocean Mixing Experiment (HOME; \citet{Rudnick-etal-2003}), full-depth profiles of horizontal velocity, temperature, salinity and pressure were collected during 2000 and 2002 \citep{Lee-etal-2005} using the Absolute Velocity Profiler (AVP; \citet{Sanford-etal-1985}). These profiles are analyzed with wavenumber bispectra to look for PSI with the M$_2$ tide.

Data are transformed into a coordinate system aligned with large-scale local bathymetry, so
that cross-ridge is 37.4$^\circ$E of N and along-ridge is 37.4$^\circ$N of W. Vertical resolution was about 10~m.

Perturbation velocities were calculated by removing the barotropic velocity and time mean as well as a residual depth-average to insure the perturbation velocities satisfy baroclinicity,
\begin{equation}
u'(z,t)=u(z,t)-\left<u(z,t)\right>_t-\frac{1}{H}\int_{-H}^0\left[u(z,t)-\left<u(z,t)\right>_t\right]\,dz\ .
\end{equation}
where here the prime denotes a perturbation velocity.
Data were WKBJ-normalized, transforming to a constant-$N$ ocean (Althaus et al., 2003). Stratifi- cation is an unresolved problem for computing energy transfer rates of PSI (See Appendix A).

Bispectra and bicoherences were computed with an ensemble average over AVP drops within a 5-km radius. The results are very noisy. Bispectra for two stations (A and B, Figure 11) sampled within days of each other during the spring phase of the spring-neap cycle are shown in Figure 12. Bottom depths are 4700 (green) and 4400 m (red). Even with these similarities, the modal content and bispectra at the two stations are very different. The bispectrum at station A (Figure 12a) is red, with some larger values near vertical wavenumber pair $(0.006,0.006)$ (about mode-5) rad m$^{-1}$. Bicoherences at the same wavenumbers are elevated. The bispectrum at station B has peaks at higher wavenumbers,                        (0.002,0.014) rad m$^{-1}$  (mode-1 and mode-10). Again, bicoherences are elevated at these same peaks. The bicoherence levels for station A are high ($\approx 0.7$). Assuming our six profiles represent six independent realizations, this would correspond to a significance level of 90\%, similar to values found by \citet{Carter-Gregg-2006}. However, it may not be reasonable to assume that six profiles sampled within 36 hours each constitute an independent realization. Additionally, absence of frequency information and limited availability of data, combined with variability between bispectra at two nearby stations preclude unambiguously identifying these signals as PSI.

Variability in the signal may have several causes.  Near the generation site, where AVP was
dropped, the M$_2$ internal tide forms beams of energy comprised of many modes radiating from the ridge. Even if the ratio of daughter-wave to parent-wave wavenumber is constant, the actual daughter-wave wavenumbers will be spread over multiple wavenumbers. If the ridge were a true two-dimensional knife-edge, modal content in the M$_2$ would be the same at the two sites sampled. Their bispectra, though peaks might be spread in wavenumber space, would have peaks at the same values for the two stations. However, complex local topography has a strong effect on direction of energy-flux, beam position, and modal composition.

The distance between the sampling location and generation site also affects characteristics of
PSI. Near the site of internal wave generation, many modes of M$_2$  energy exist.  Higher modes dissipate near the ridge, a process which may be aided by PSI \citep{Carter-Gregg-2006}. A few hundred kilometers from the ridge, only the mode-1 internal tide continues to propagate. If it is unstable to PSI, the frequency and wavenumber bispectra may be used to determine rates of energy transfer. The model simulates the latter case.

Temporal variability is a lesser concern, since the M$_2$ barotropic tidal signal is relatively stationary and the ridge undergoes changes on a much slower timescale than the waves. But phase of the spring-neap cycle when data are collected greatly affects the strength of the semidiurnal tide, and thus the strength of the PSI interaction. Another concern is addressed in \citet{MacKinnon-Winters-2005}, that PSI dominates near the so-called `turning latitude' where $0.5\omega_{M_2}=2f$ or 28.9$^\circ$, but is much weaker at lower latitudes (AVP data were from 21.6$^\circ$N).

An ideal dataset would have both frequency and wavenumber information, as from a profiling
mooring, fast CTD or ADCP of velocity and density, in a region of strong M$_2$ tide with constant mode number below 28.9$^\circ$ latitude.  With a flat bottom and weak background, the bispectrum may provide realistic estimates of energy transfer through the vertical wavenumber spectrum via parametric subharmonic instability.

\section{Conclusions}

The motivation for this work is to better-understand the role of wave-triad interactions in the cascade to turbulence. To this end, we explored applications of the bispectrum and bicoherence for identifying wave-triads and quantifying the energy transferred through the internal wave spectrum. Emphasis was given to tailoring the tools for PSI of a coherent internal tide and demonstrating the quantification of energy transfer.

Identification of PSI of a coherent internal tide has a couple steps.  In frequency space, the signature of PSI is a triad of three waves, two with half the frequency of the third. The bispectrum of a model run with a mode-1, M$_2$ internal forcing at 21$^\circ$N latitude in Figure 7. In wavenumber space, it is helpful to begin by determining the expected wavenumber signature of PSI in the bispectrum. For a given latitude and departure of daughter waves from exactly subharmonic PSI, ellipse (8) is the locus of wavenumbers satisfying the resonance conditions for PSI, plotted in Figure 4 as a function of departure from exactly subharmonic.  The wavenumber bispectrum of the same model run (Figure 8) has peaks at the predicted ellipse, verifying that the peaks are PSI, and in both bispectrum and bicoherence, confirming that the triad is energetic and phase-locked. Phase-locking is a requirement for energy transfer.

The transfer bispectrum (26) estimates the energy evolution using bispectra. Energy transfer
rates are computed for a spinup period of a model run and compared to energy buildup in the subharmonic wave (Figure 10). The slope and amplitude agree well, confirming the bispectrum?s ability to estimate energy transfer rates. Computing energy transfer rates via PSI is the best way to determine whether the interaction is physically relevant.  Energy transfer rates depend on a combination of high energy levels and phase-locking, and one may be compensated for by the other. That is to say, even when bicoherence values indicate a lower-degree of phase-locking triad waves, if the triad is energetic enough, energy may still transferred. If multiple dimensions of data are not available, there are two ways to identify PSI. In frequency space, peaks in bispectrum and bicoherence at half-frequency triads confirm the existence of PSI. In wavenumber space, peaks at in the bispectrum bicoherence at wavenumber solutions to ellipse 8 for an independently determined $\epsilon$
indicate PSI is occurring.

The transfer bispectrum also identifies which nonlinear terms are primarily responsible for energy transfer. In the case of the model, we find that the $u\partial u/\partial x$ term dominates transfer. The importance of this term is contrary to previous suggestions of the buoyancy terms $w\partial b/\partial z$ or $u\partial b/\partial x$.

An attempt is made to use the wavenumber bispectrum to identity PSI in ocean profiles of horizontal velocity, but the signal was inconclusive (Figure 12). The effects on bispectra and on PSI of tidal beams, broadband M$_2$ energy in mode-space, complex topography, temporal variability and latitudinal dependence of the interaction were discussed. In regions, such as near the Hawaiian Ridge, analysis or multi-dimensional data, which are less likely to be contaminated by the aforementioned issues, may be the only way to reliably identify PSI triads.

Several open questions follow from this work. Though triad ellipses are constructed in terms of $\epsilon$, a small parameter indicating how far daughter-wave frequencies are from exactly subharmonic, what sets $\epsilon$? If it is set by external forces and could be computed without frequency information, then the triad ellipses could be used to make the wavenumber bispectra nearly as useful as frequency bispectra for identifying PSI. Energy transfer by the bispectrum depends on wavenumber, amplitude and degree of phase-locking of triad waves.  The steady-state rates of energy transfer set levels of energy dissipation in the ocean internal wave spectrum. What then is the rate-limiting step in the cascade of energy through PSI: the energy source, PSI or loss of energy from daughter waves via other wave-triad interactions? Future studies involving varying levels or wavenumbers of the source wave may shed light on these questions.

\appendix
\section{Stratification}

The model used to simulate PSI has constant buoyancy frequency, which simplifies the physics and computation of energy transfer via the bispectrum. In the ocean, perturbation data scale with stratification, so that horizontal velocities are higher in strongly stratified than weakly stratified waters (Gill, 1982). Since PSI transfer rates are proportional to energy in the waves, transfer via the $u\partial u/\partial x$ term would be higher in stratified areas than non-stratified areas. Furthermore, computation of the transfer bispectrum is complicated by stratification since WKBJ-normalizing data changes the scaling of profiles, eliminating the stratification dependence by changing the magnitude of transfer rate estimates. One way to combat this would be to use energy transfer estimates computed separately for depth bins of different stratification. This would suffice if triad waves were all of high wavenumber and essentially confined to small regions of constant stratification. But since lower-mode waves propagate further from sources without dissipation, mode-1 may be the primary M$_2$ parent wave in a PSI interaction far from generation sites.

\section{Isopycnal  Coordinates}

\citet{Neshyba-Sobey-1975} used bispectra to identify PSI in Arctic Ocean data, but the signal was later discovered to be not due to PSI but rather from kinematic contamination or Doppler-shifting of a waves? Eulerian frequency by a background mean flow or other waves. This process is known to create harmonics at sums of fundamental frequencies, and may be difficult to distinguish from a true bispectral signal \citep{McComas-Briscoe-1980,Williams-1985}.  Transforming data into isopycnal coordinates reduces total frequency Doppler-shifting of daughter waves, which may be quantified,
\begin{equation}
\frac{\Delta \omega'}{\omega'}=\frac{k_1u+k_2v+k_3w}{\omega'}
\end{equation}
(Gill, 1982).   The effect of Doppler shifting in the model is approximately $\Delta\omega'/\omega'\approx 0.05$. Transforming into isopycnal coordinates eliminates the effect of $k_3\omega$ in the Doppler shift.  Us- ing multi-dimensional bispectra may also avoid this problem, since Doppler-shifted waves do not have wavenumbers which satisfy wavenumber resonance conditions and the dispersion relation.

\begin{figure}
\centering\includegraphics[width=0.5\textwidth]{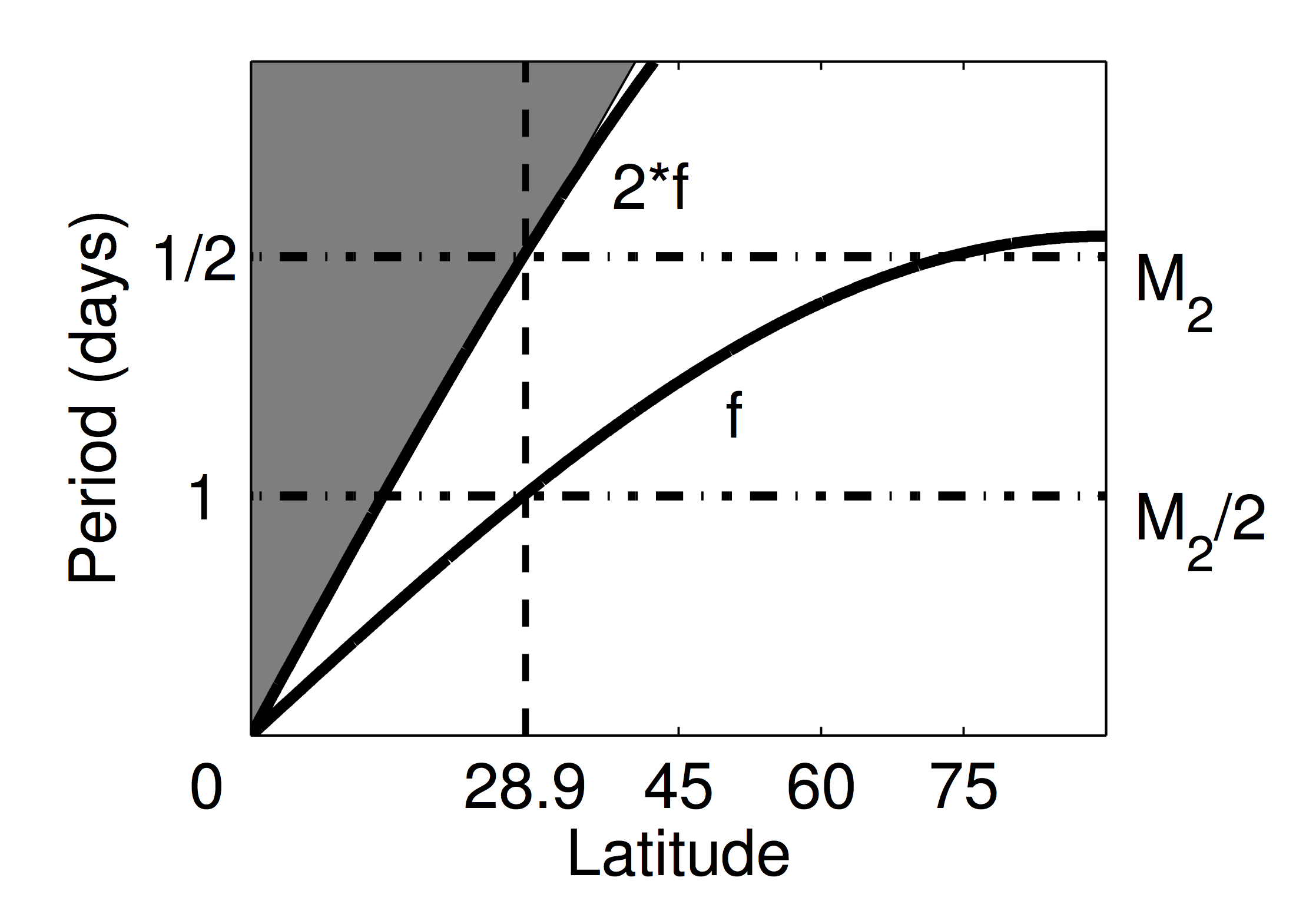}
\caption{Frequency restrictions on waves in a PSI triad as a function of latitude.  The shaded region shows frequencies allowable for the parent wave of a PSI triad which are greater than twice the local Coriolis frequency. PSI of the M$_2$  tide may occur equatorward of $2f=\omega_{M_2}$ or 28.9$^\circ$.}
\end{figure}

\begin{figure}
\centering\includegraphics[width=\textwidth]{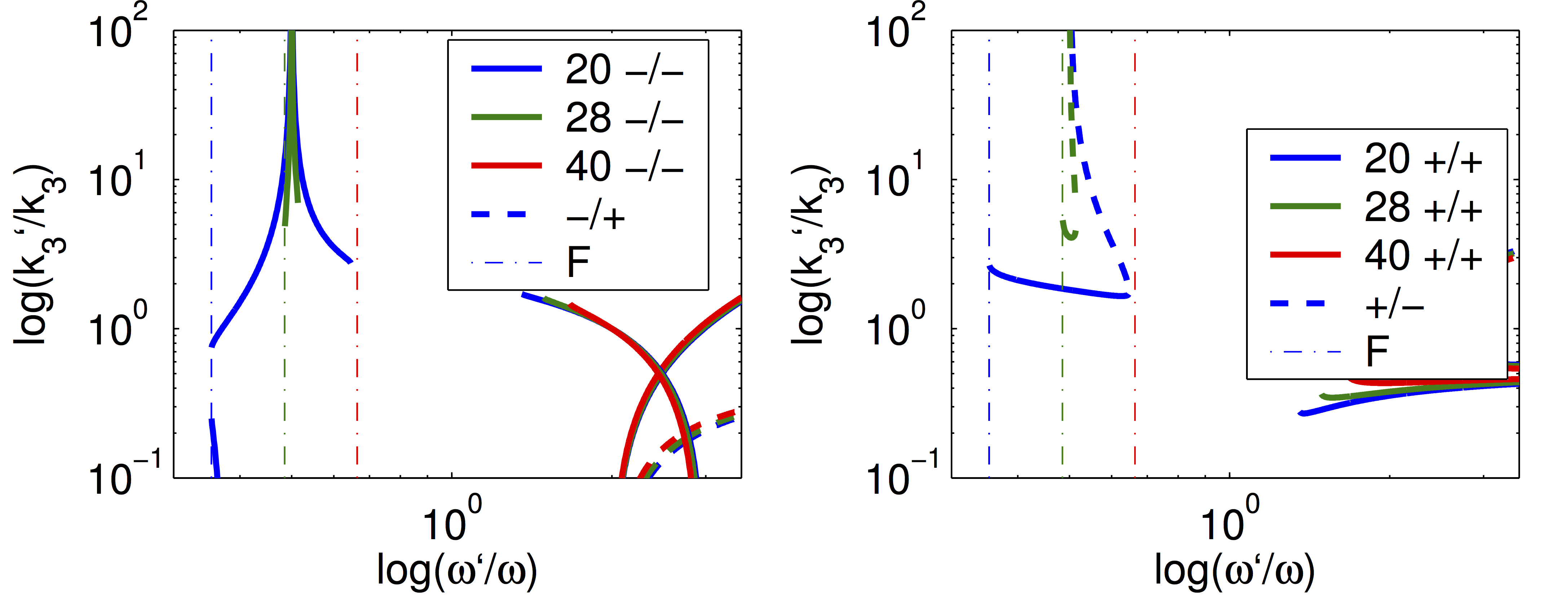}
\caption{General triad solutions to the resonance conditions and dispersion relation (5).  The left plot shows solutions for a minus sign choice in the numerator and a minus (plus) choice in the denominator as a solid (dashed) line.  The right plot shows solutions for a plus sign in the numerator and a plus (minus) sign in the denominator as a solid (dashed) line. Solutions are shown for latitudes 20, 28 and 40, as a function of frequency and vertical wavenumber where $\omega$ is the M$_2$  frequency. PSI is characterized by $\omega'/\omega\approx 1/2$, which is near the green vertical dashed line on the plots. Wavenumber-frequency solutions for PSI at each latitude are plotted in bold. Higher wavenumber solutions are called induced diffusion, an interaction which smooths energy spikes in wavenumber spectra to nearby wavenumbers. The dashed lines are the local Coriolis frequency for each wavenumber, which show minimum frequencies for an internal wave. Near latitude 28, the Coriolis frequency is approximately half the M$_2$  frequency. The minus sign allows higher PSI daughter wavenumbers than the plus sign for the same frequency.}
\end{figure}

\begin{figure}
\centering\includegraphics[width=0.5\textwidth]{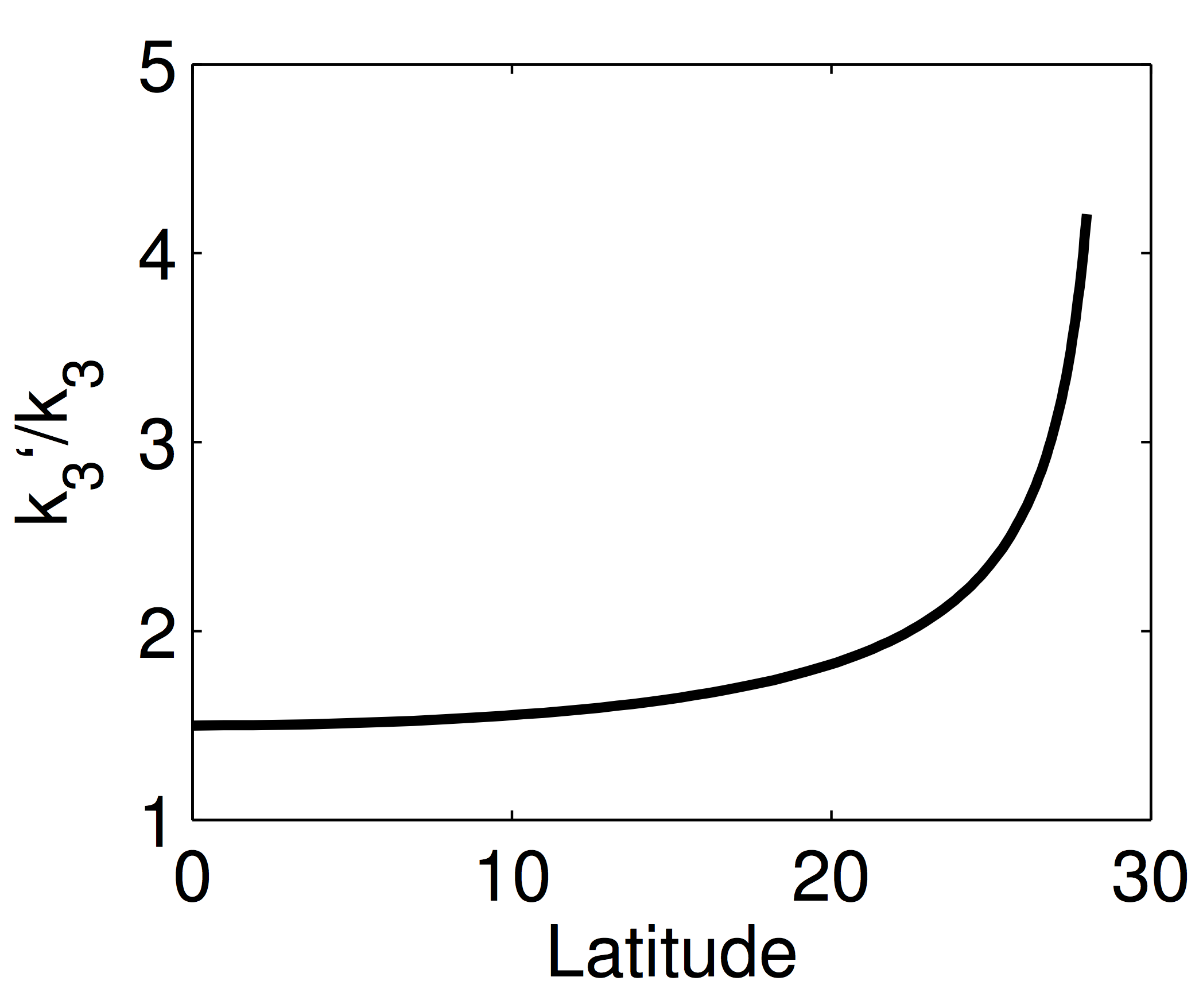}
\caption{Solutions to the resonance conditions and dispersion relation for exactly subharmonic
PSI with the M$_2$  semidiurnal tide as the parent wave, $\omega'=\omega''=\omega/2$ as a function of latitude.
}
\end{figure}

\begin{figure}
\centering\includegraphics[width=0.5\textwidth]{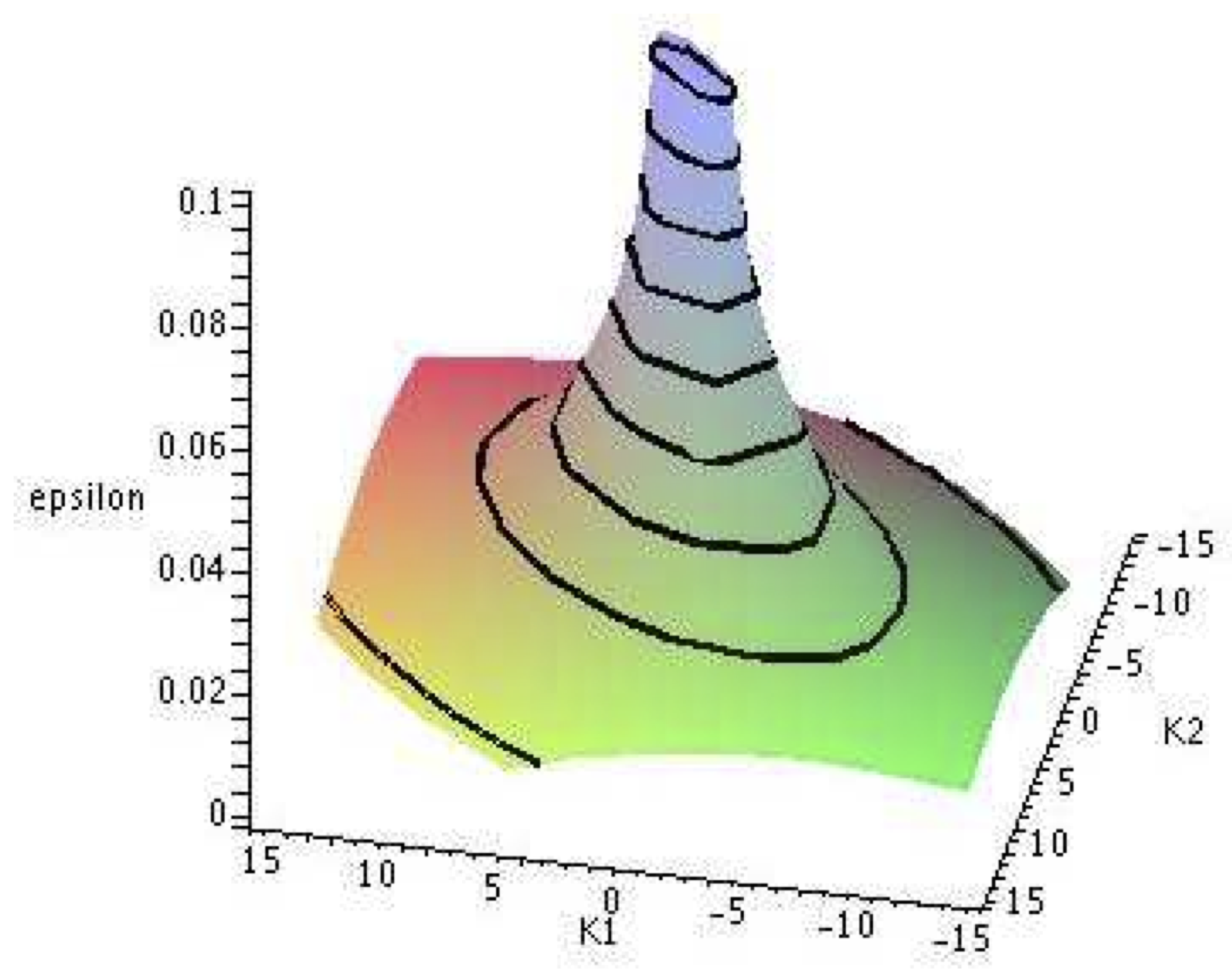}
\caption{Nondimensional vertical wavenumber ellipse as a function of $\epsilon$, the deviation of the daughter-wave frequencies from half the parent-wave frequency, at 21$^\circ$ latitude from (8). As $\epsilon\longrightarrow0$, the ellipse size increases, or the ratio of daughter to parent vertical wavenumber increases.
}
\end{figure}

\begin{figure}
\centering\includegraphics[width=\textwidth]{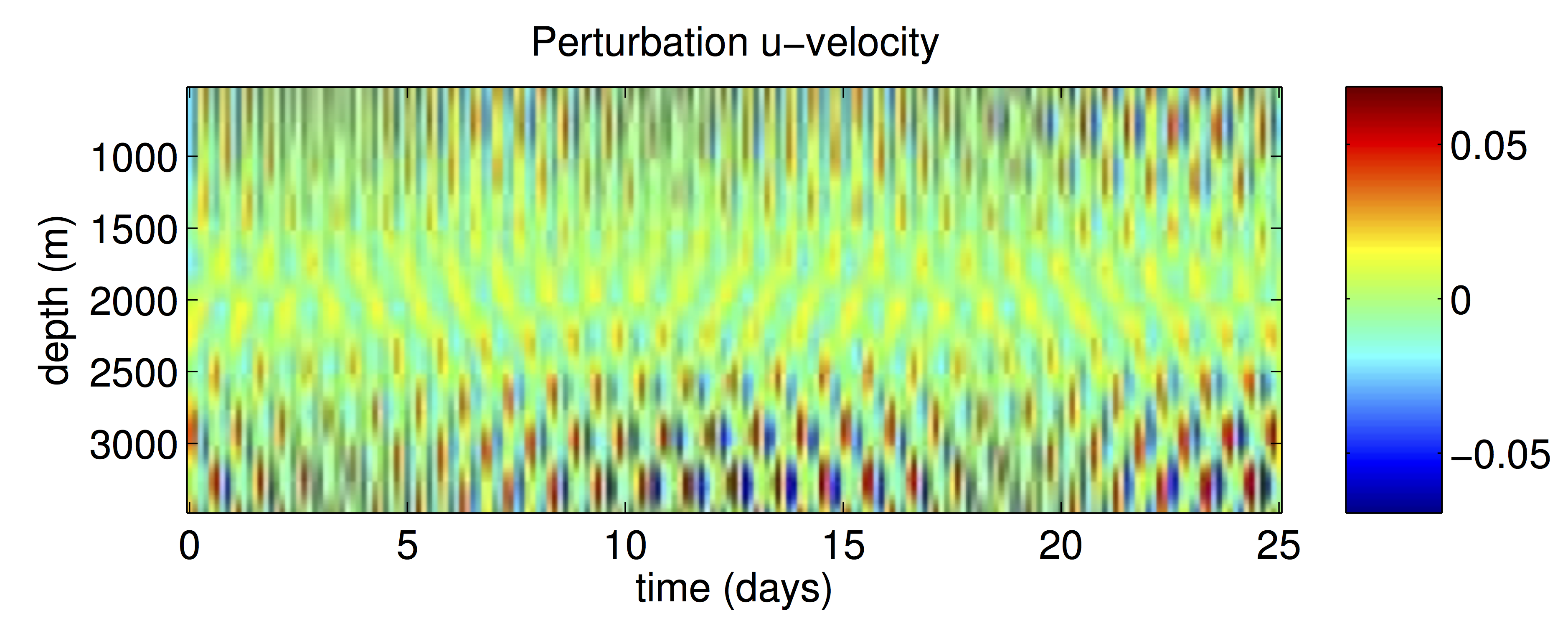}
\caption{Perturbation  $u$-velocity profile time-series from model run in steady state. Though forced by a mode-1 M$_2$ semidiurnal tide, higher wavenumbers are clearly visible near the top and bottom of the ocean.  In the numerical model, the location of onset of PSI varies due to random noise inserted in the model.}
\end{figure}

\begin{figure}
\centering\includegraphics[width=\textwidth]{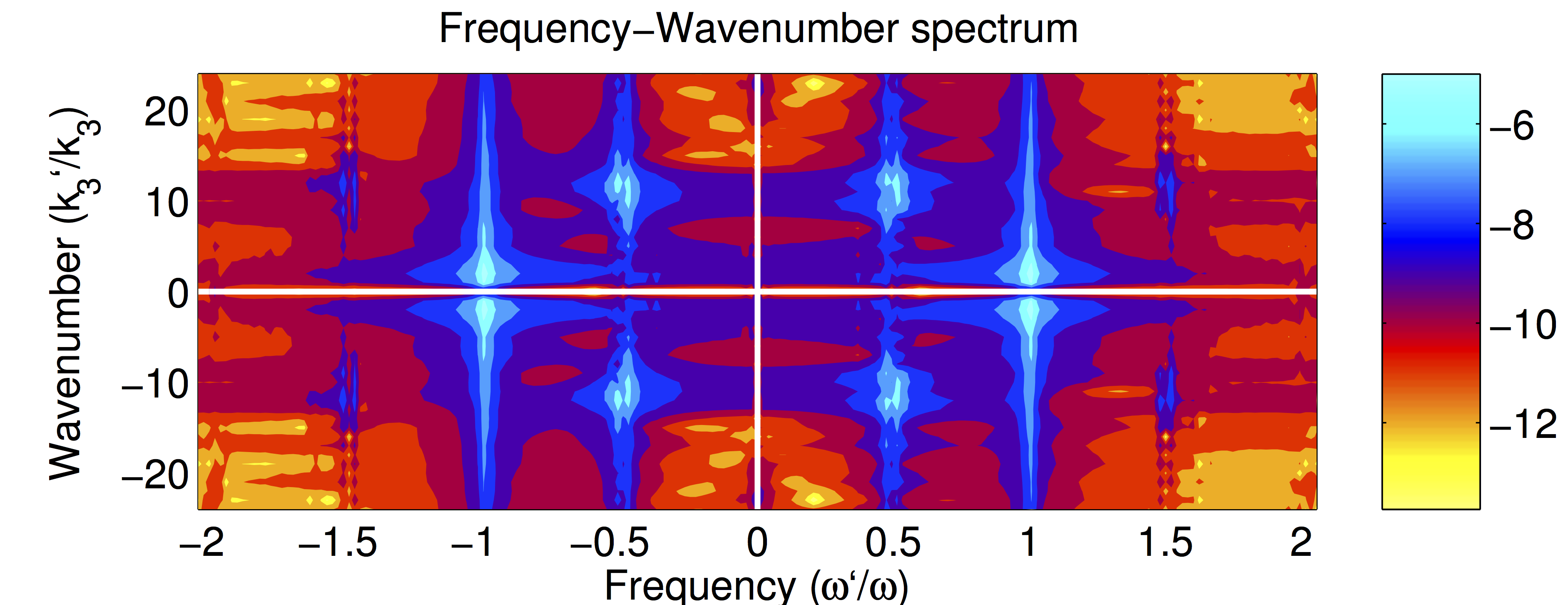}
\caption{Frequency-vertical wavenumber power  spectrum of$u$ velocity data,  in  units  of $\log_{10}(\mbox{m}^2/\mbox{s}^2)$.  Frequency is normalized by the M$_2$  frequency; vertical wavenumber by mode-1 wavenumber.  Since the only energy source in the model is given by the peak at $(\omega,k_3)=(\pm1,\pm1)$, the peaks at $(\pm0.5,\pm10)$ indicate energy has been transferred to this frequency-wavenumber band by some nonlinear process.}
\end{figure}

\begin{figure}
\centering\includegraphics[width=.5\textwidth]{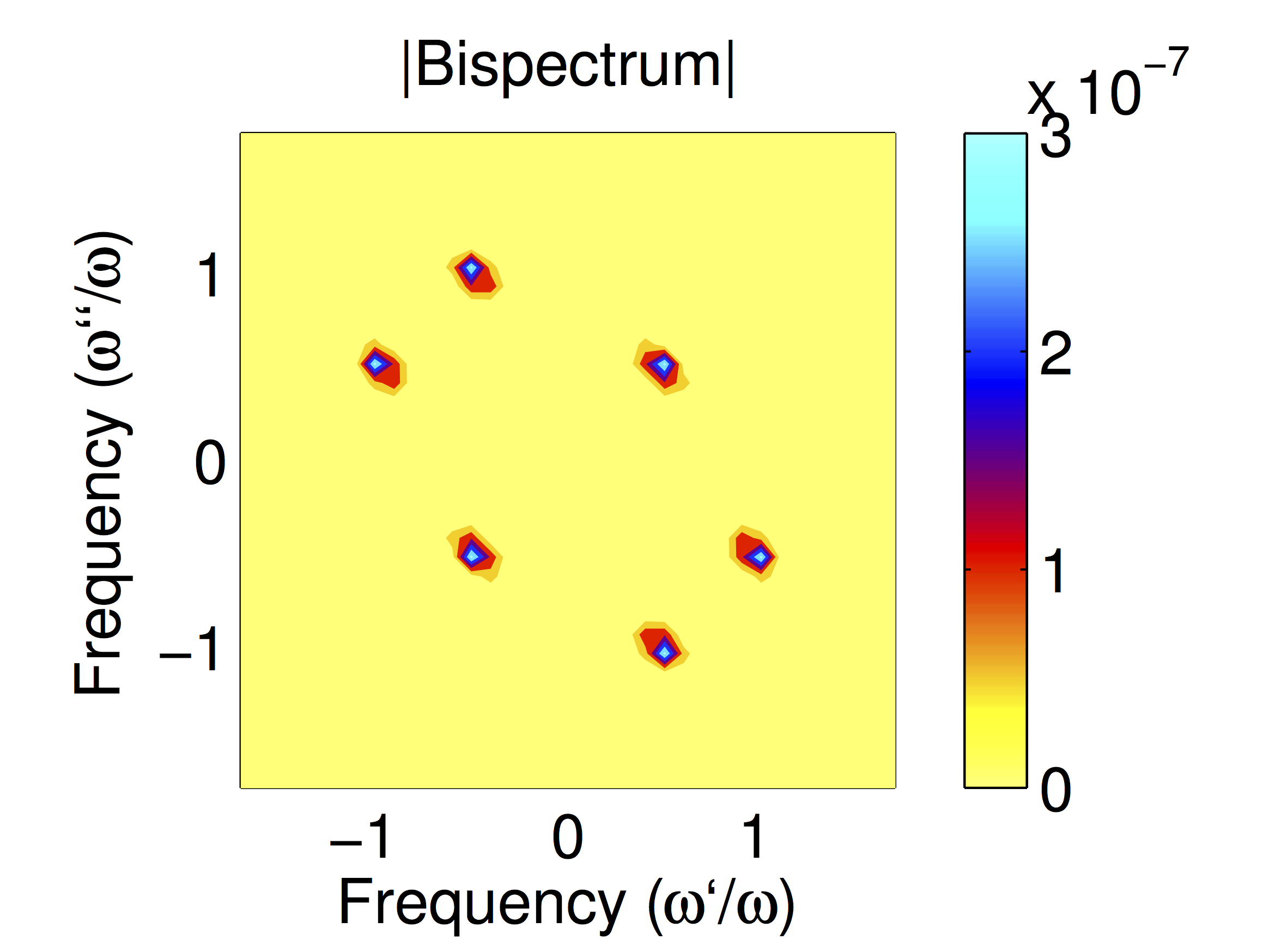}
\caption{Frequency bispectrum (m$^3$ s$^{-3}$)  of model data. Axes are normalized by the M$_2$  frequency. The peak at $(\omega'/\omega,\omega''/\omega) = (1/2,1/2)$  corresponds to the energetic triad $(0.5\omega_{M_2},0.5\omega_{M_2},\omega)$. Additional peaks are due to symmetries in the bispectrum.}
\end{figure}

\begin{figure}
\centering\includegraphics[width=\textwidth]{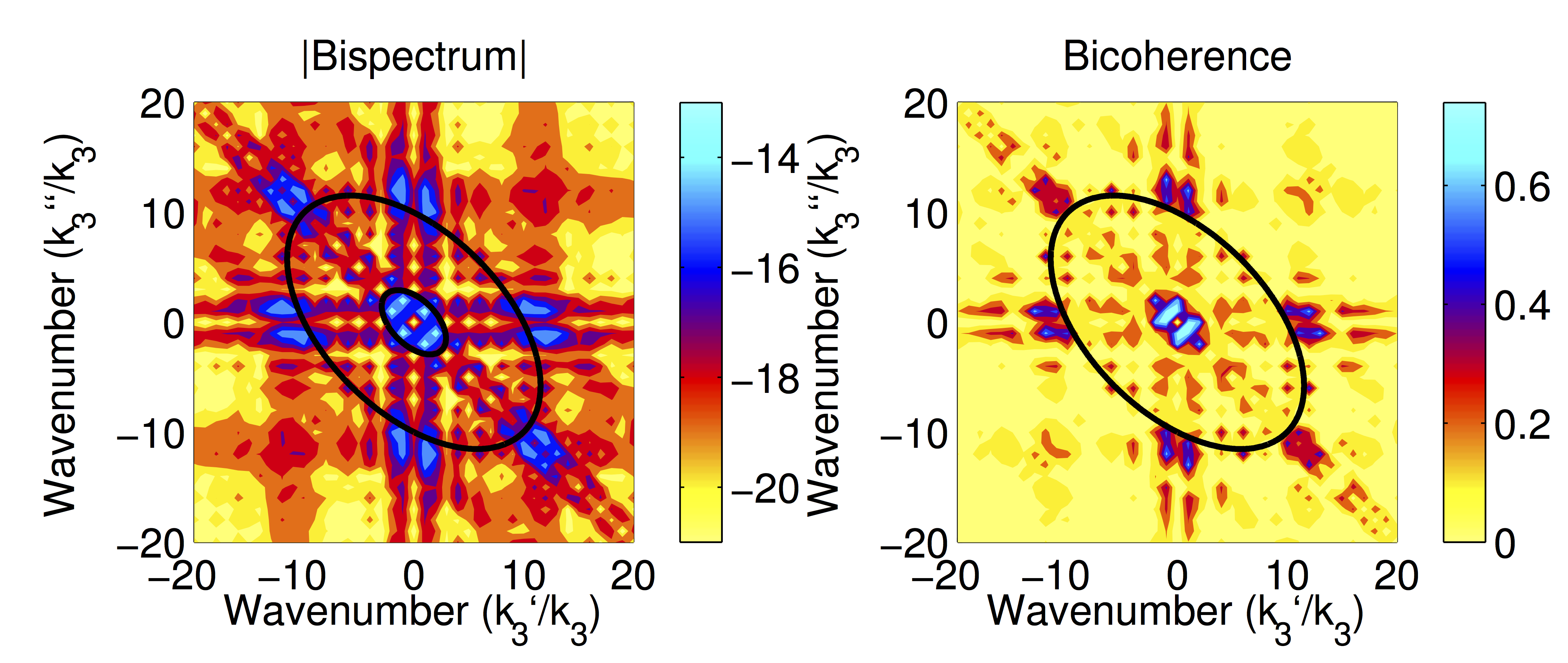}
\caption{(a) Vertical wavenumber bispectrum and (b) bicoherence of model data. Black ellipses are the theoretical patterns predicted for $\epsilon=0.02$. This $\epsilon=0.02$ can also be seen in frequency spectra for the same data. Axes are normalized by the mode-1 wavenumber.}
\end{figure}

\begin{figure}
\centering\includegraphics[width=\textwidth]{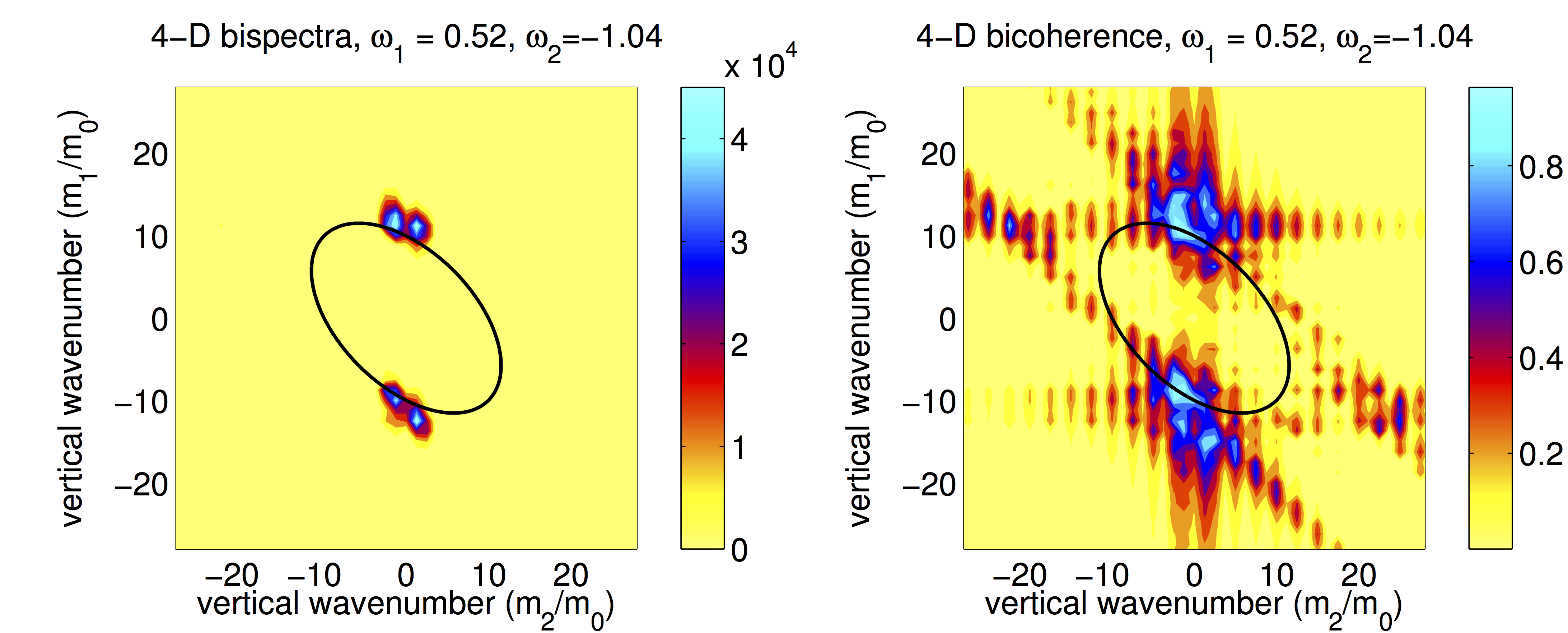}
\caption{Wavenumber bispectrum (a) and bicoherence (b) from the 4-D bispectrum selected for the 0.5M$_2$  and $-$M$_2$  frequencies.  The black ellipse is the theoretically predicted nondimensional wavenumber ellipse with  $\epsilon=0.06$.}
\end{figure}

\begin{figure}
\centering\includegraphics[width=.5\textwidth]{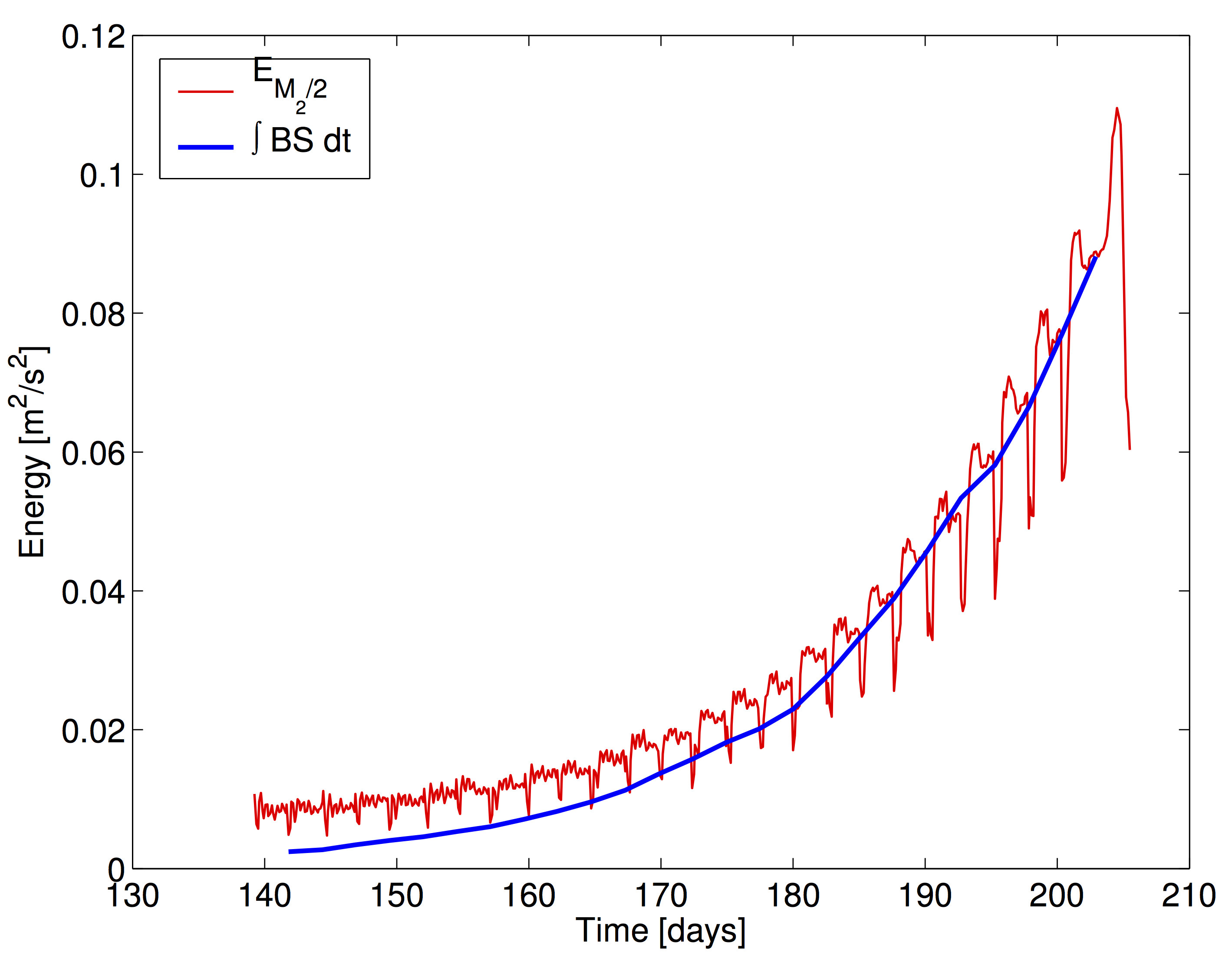}
\caption{A comparison between energy buildup in the subharmonic (red) and the time-integrated energy transfer into the subharmonic with PSI as computed via the bispectrum (blue).  Both are averages over the   $x$-direction. The integrated bispectrum is effectively a running average of energy transferred. Dissipation and loss of energy to other triads are not accounted for in the blue curve.
 }
\end{figure}

\begin{figure}
\centering\includegraphics[width=.5\textwidth]{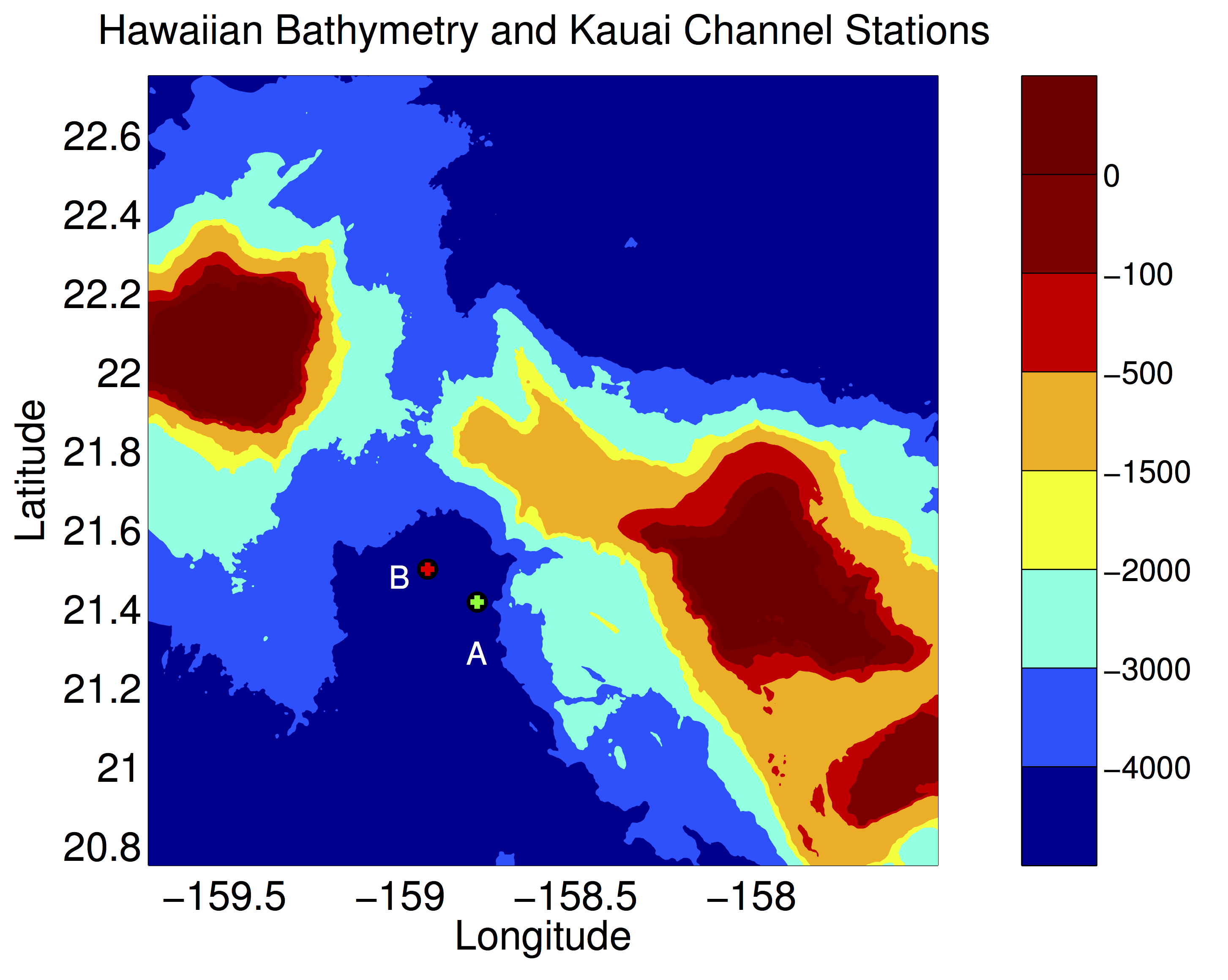}
\caption{Bathymetry of the Hawaiian Ridge between Oahu and Kauai.  Stations analyzed are marked in green and red.}
\end{figure}

\begin{figure}
\centering\includegraphics[width=\textwidth]{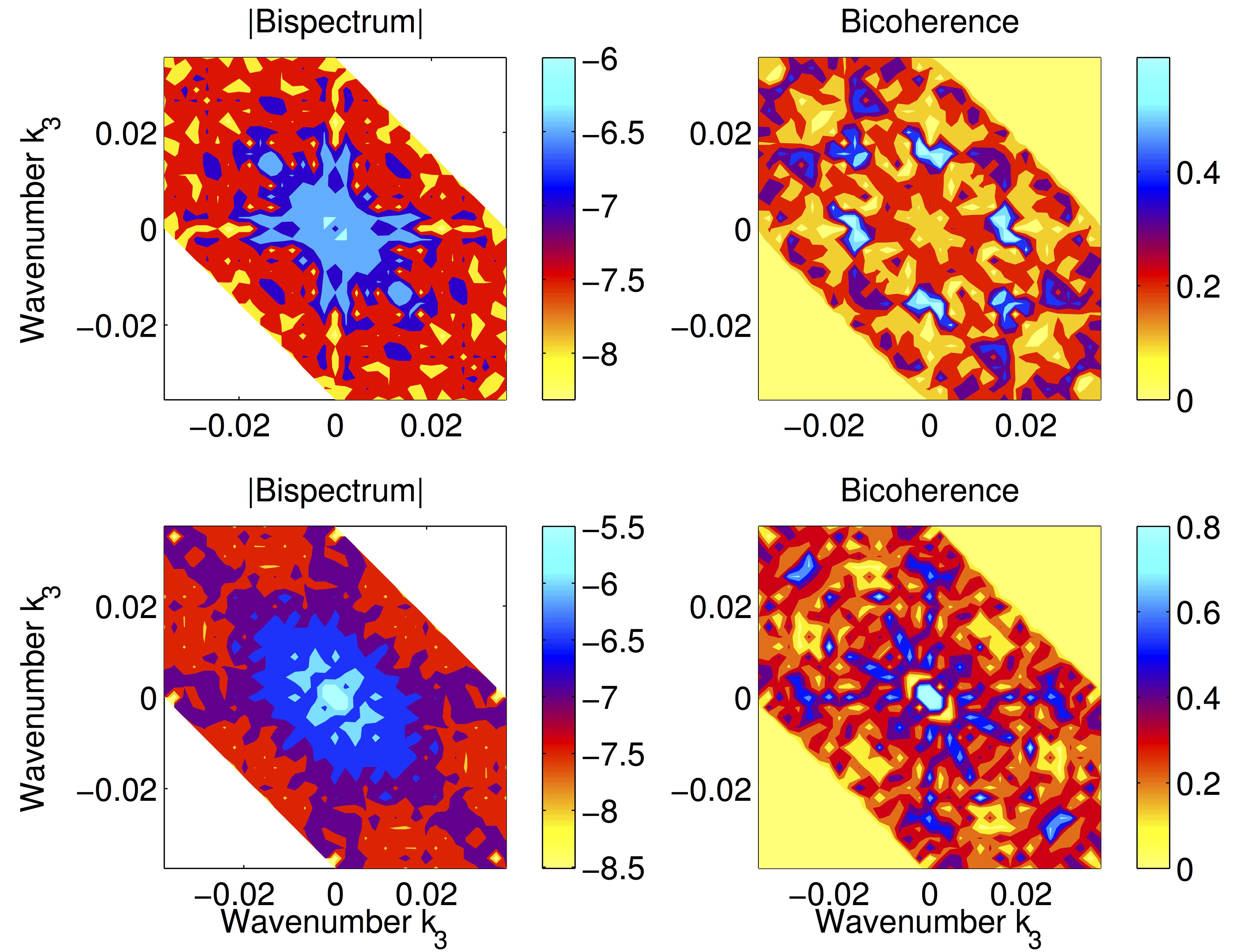}
\caption{Bispectra (a and c) and bicoherences (b and d) for WKBJ-normalized across-ridge perturbation velocities. Panels (a) and (b) are for the deeper station marked in green in Figure 11, and (c) and (d) for the station marked in red.}
\end{figure}

\clearpage
\begin{small}

\end{small}
\end{document}